\documentclass[%
 amsmath, amssymb, nofootinbib,
 reprint, superscriptaddress%
]{revtex4-2}

\usepackage[utf8]{inputenc}

\usepackage{lipsum}
\usepackage{graphicx}
\usepackage{dcolumn}
\usepackage{bm}
\usepackage[T1]{fontenc}
\usepackage{mathptmx}
\usepackage{braket}
\usepackage{booktabs}
\usepackage{hyperref}
\hypersetup{colorlinks=true,linkcolor=blue,citecolor=blue,filecolor=blue,urlcolor=blue}
\usepackage{listings}
\usepackage{microtype}

\newcommand{\autoqmlshort}{10\, 000}
\newcommand{\autoqmllong}{50\, 000}

\begin{document}

\title{AutoQML: A Framework for Automated Quantum Machine Learning} 

\author{Marco Roth}
\email{marco.roth@ipa.fraunhofer.de}
\affiliation{Fraunhofer Institute for Manufacturing Engineering and Automation IPA, Nobelstraße 12, 70569 Stuttgart, Germany}

\author{David A. Kreplin}
\email{david.kreplin@ipa.fraunhofer.de}
\affiliation{Fraunhofer Institute for Manufacturing Engineering and Automation IPA, Nobelstraße 12, 70569 Stuttgart, Germany}

\author{Daniel Basilewitsch}
\affiliation{TRUMPF SE + Co. KG, Quantum Applications Group, Johann-Maus-Straße 2, 71254 Ditzingen, Germany}

\author{João F. Bravo}
\affiliation{Fraunhofer Institute for Industrial Engineering IAO, Nobelstraße 12, 70569 Stuttgart, Germany}

\author{Dennis Klau}
\affiliation{Fraunhofer Institute for Industrial Engineering IAO, Nobelstraße 12, 70569 Stuttgart, Germany}

\author{Milan Marinov}
\affiliation{USU GmbH, R\"uppurrer Str. 1, 76137 Karlsruhe, Germany}

\author{Daniel Pranji\'{c}} 
\affiliation{Fraunhofer Institute for Industrial Engineering IAO, Nobelstraße 12, 70569 Stuttgart, Germany}

\author{Horst Stuehler}
\affiliation{Zeppelin GmbH, Graf-Zeppelin-Platz 1, 85748 Garching, Germany}

\author{Moritz Willmann}
\affiliation{Fraunhofer Institute for Manufacturing Engineering and Automation IPA, Nobelstraße 12, 70569 Stuttgart, Germany}

\author{Marc-Andr\'e Z\"oller}
\affiliation{USU GmbH, R\"uppurrer Str. 1, 76137 Karlsruhe, Germany}

\date{\today}

\begin{abstract}
Automated Machine Learning (AutoML) has significantly advanced the efficiency of ML-focused software development by automating hyperparameter optimization and pipeline construction, reducing the need for manual intervention. Quantum Machine Learning (QML) offers the potential to surpass classical machine learning (ML) capabilities by utilizing quantum computing. However, the complexity of QML presents substantial entry barriers. We introduce \emph{AutoQML}, a novel framework that adapts the AutoML approach to QML, providing a modular and unified programming interface to facilitate the development of QML pipelines. AutoQML leverages the QML library sQUlearn to support a variety of QML algorithms. The framework is capable of constructing end-to-end pipelines for supervised learning tasks, ensuring accessibility and efficacy. 
We evaluate AutoQML across four industrial use cases, demonstrating its ability to generate high-performing QML pipelines that are competitive with both classical ML models and manually crafted quantum solutions.
\end{abstract}

\maketitle

\section{Introduction}

A key factor in the success and democratization of machine learning (ML) has been the development of increased abstraction levels, which facilitate rapid prototyping and lower entry barriers. Automated machine learning (AutoML) aims to automate the predominantly manual process of ML pipeline construction, representing a significant progression in this development~\cite{zoeller_2021b}. This approach has proven successful by enhancing the efficiency of specialists, allowing them to focus more on modeling business problems rather than on implementation details~\cite{Wang2021}. Furthermore, the reduced expertise required to use these tools democratizes ML methods, allowing companies with less experience to integrate ML-based solutions into their workflows~\cite{hutter2019automated}. This is a particularly relevant consideration in market environments that are increasingly impacted by labor shortages, notably in specialized domains such as ML~\cite{Miller2017}.

Quantum machine learning (QML) employs quantum computers to develop ML algorithms that harness quantum mechanical principles, with the goal of expanding the capabilities of ML beyond the classical limits~\cite{Liu2021, doi:10.1126/science.abn7293, Liu2024}. As an interdisciplinary field requiring expertise in quantum computing, ML, and computer science, the entry barriers to this technology are particularly high. In this work, we introduce the framework \emph{AutoQML}\footnote{Code available at \href{https://github.com/AutoQML/autoqml}{https://github.com/AutoQML/autoqml}}, which aims to transfer the success of AutoML from the classical to the quantum realm, making QML accessible to a broad audience in science, technology and industry.

Using quantum computers for ML instead of classic hardware presents a unique set of challenges. Providing a diverse suite of QML algorithms that can be orchestrated using AutoML techniques, such as combined algorithm selection and hyperparameter optimization (CASH)~\cite{10.1145/2487575.2487629} requires a modular implementation with a unified programming interface. Additionally, a seamless transition from classical preprocessing and simulation to real quantum computers needs to be ensured. AutoQML builds on the QML library sQUlearn~\cite{Kreplin2025}, which offers a variety of QML algorithms with a scikit-learn~\cite{scikit-learn} programming interface to create a set of modular and diverse QML methods. The library leverages PennyLane~\cite{bergholm2022pennylane} and Qiskit~\cite{QiskitCommunity2017}, enabling the execution of algorithms on multiple simulators and quantum computers, such as IBM Quantum~\cite{IBMQuantum} and various backends available through Amazon Braket~\cite{braket}.

AutoQML creates end-to-end QML pipelines for various supervised learning scenarios, such as time series classification, tabular regression, and image classification. Using the open-source libraries Optuna~\cite{Akiba2019} and Ray Tune~\cite{liaw2018tune}, the framework offers fully optimizable pipelines, including quantum-specific preprocessing, to make QML accessible to non-experts. In designing AutoQML, we have anticipated future developments in quantum computing and focused particularly on modularity, allowing for easy extension of the algorithm pool.

In this work, we outline the architecture of the AutoQML framework. We benchmark the framework on four distinct industrial use cases involving time series and image classification, as well as tabular and time-series regression. The results are compared to classical solutions and manual quantum computing pipelines. This study aims to demonstrate the capability of AutoQML in facilitating the development of effective QML solutions in various domains.

\begin{figure*}[tb]
    \centering
    \includegraphics[width=0.7\linewidth]{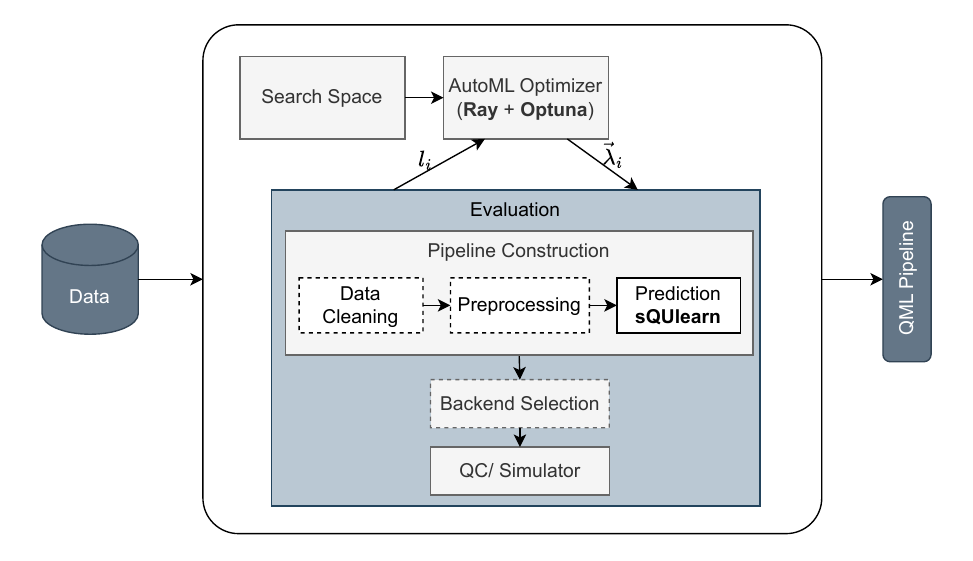}
    \caption{Architecture overview of the AutoQML framework. Data is supplied by the user. Using Ray and Optuna, AutoQML constructs a pipeline that is optimized over a preconfigured search space. A loss value $l_i$ is obtained for each configuration $\vec{\lambda}_i$, which consists of data cleaning, preprocessing, and a model with hyperparameters evaluated using a simulator or a real quantum computer (QC). After a given budget is exhausted, the best-performing pipeline is returned to the user. Optional pipeline steps are indicated as dashed boxes.}
    \label{fig:autoqml-lib-architecture}
\end{figure*}

The remainder of this work is structured as follows. Following the related work in Sec.~\ref{sec:related_work}, Sec.~\ref{sec:framework} outlines the architecture of the AutoQML framework. Section~\ref{sec:use_cases} describes the industrial use cases to evaluate the performance of the framework. Section~\ref{sec:results} presents the experimental results, comparing AutoQML-generated pipelines with both manually crafted QML pipelines and classical ML approaches. Finally, Secs.~\ref{sec:conclusion} and \ref{sec:conclusion} conclude with a summary and discussion of our findings and potential future research directions.

\subsection{Related Work}
\label{sec:related_work}

The term AutoQML has been used previously, primarily emphasizing the optimization of specific QML models rather than addressing the optimization of the complete pipeline, which includes preprocessing and the selection from different QML models.
This term has been applied to hyperparameter optimization for a fixed model~\cite{Gomez2022}, as well as in the context of architecture search for optimizing the encoding of data into quantum states~\cite{Altareslopez2024, KoikeAkino2022}.
In these architecture search efforts, the QML model remains fixed while the encoding circuit is optimized for the dataset, aiming to improve the performance of the given model~\cite{Altares-López_2021, Incudini2024, Dai2024, Rapp_2025}.
Our work distinguishes itself from these prior approaches by presenting, to the best of our knowledge, the first comprehensive framework that constructs and optimizes end-to-end QML pipelines. Unlike previous efforts that focused on optimizing specific QML models or encoding circuits, our approach emphasizes the integration of diverse QML methods, enabling the development of adaptable and effective QML solutions tailored to various application scenarios.

\section{Framework}
\label{sec:framework}

The foundation of AutoQML are the open-source optimization, parallelization, and scheduling frameworks Ray Tune and Optuna. The algorithm pool is constituted by scikit-learn and sQUlearn, with extended functionalities designed to optimize and streamline the development of QML pipelines. Neural network-based approaches such as autoencoders for preprocessing are implemented using PyTorch~\cite{10.5555/3454287.3455008}.

\subsection{Pipeline Creation}
The framework follows a standard AutoML approach known as \emph{pipeline synthesis and optimization}~\cite{zoeller_2021a, zoeller_2021b}, designed to create complete (quantum) ML pipelines. This  involves searching for the optimal pipeline structure and performing CASH  optimization. For a given problem type such as tabular regression, AutoQML constructs pipelines based on best-practice templates, where the sequence of steps is predefined but the specific methods an. Each pipeline explores various data cleaning and preprocessing steps, along with testing available models and their hyperparameters. The available templates can be easily extended to accommodate additional scenarios.

Figure~\ref{fig:autoqml-lib-architecture} provides a high-level overview of the AutoQML architecture. It is primarily inspired by state-of-the-art AutoML approaches, such as those introduced in~\citet{hutter_2019} and \citet{Feurer2015}, and adheres to the design principles of scikit-learn~\cite{sklearn_api} to provide a well-known, standardized programming interface. Users are required to provide an input dataset and choose a suitable problem and data type, such as tabular regression.\footnote{While an extension to other learning domains, like unsupervised learning, is possible, the focus of this work is on supervised learning.} Similar to AutoML for non-quantum ML, the AutoQML framework samples (quantum) ML pipelines from a pre-defined search space for automated optimization. This search space includes the necessary steps to create an end-to-end pipeline for (quantum) ML predictions, which can be categorized into three steps:

\begin{enumerate}
    \item \emph{Data Cleaning} is responsible for eliminating potential defects from the input data. It includes imputation of missing values, outlier removal, and encoding of categorical features.
    \item \emph{Preprocessing} transforms the data into a format suitable for quantum computing. This involves dimensionality reduction, down-sampling, feature-centric and rescaling.
    \item \emph{Prediction} uses classification or regression algorithms to generate the actual predictions. The available QML algorithms are implemented via sQUlearn. In addition, a subset of classical ML methods are optionally available. 
\end{enumerate}
Most steps in the pipeline are facultative and can be combined in various ways to create diverse of ML pipeline configurations.

The set-up for the optimization is as follows. Given a complete search space description, the optimizer iteratively draws new test configurations \(i\), denoted as \(\vec{\lambda}_i \in \Lambda\), where $\Lambda$ is a configuration space. A configuration encapsulates an abstract specification of a particular (quantum) ML pipeline. Following the CASH optimization procedure, each configuration $\vec{\lambda}_i$ jointly describes algorithms for (quantum) ML models, data cleaning, preprocessing, and prediction steps, as well as their associated hyperparameters. More specifically, for a given data set $\mathcal{D}=(X,y)$, with features $X=(\vec{x}_1,\dots,\vec{x}_N)$ from some feature space $\mathcal{X}\ni\vec{x}_j$ and labels $y\subset\mathbb{R}$, we solve the following optimization problem~\cite{zoeller_2021b}
\begin{equation}
    \vec{\lambda}^*= \text{arg min}_{\vec{\lambda}\in\Lambda} l(\mathcal{D},\vec{\lambda})\,.
    \label{eq:optimal_pipeline}
\end{equation}
Here, $l$ is a suitable loss function. AutoQML thus not only optimizes over quantum models and their hyperparameters but also over the type of preprocessing algorithms (e.g. PCA) and their configuration (e.g. the number of retained principal components).  Pipeline examples are shown in Tab.~\ref{tab:autoqml_pipeline}.

\begin{figure}[t]
    \centering
    \includegraphics[width=\linewidth]{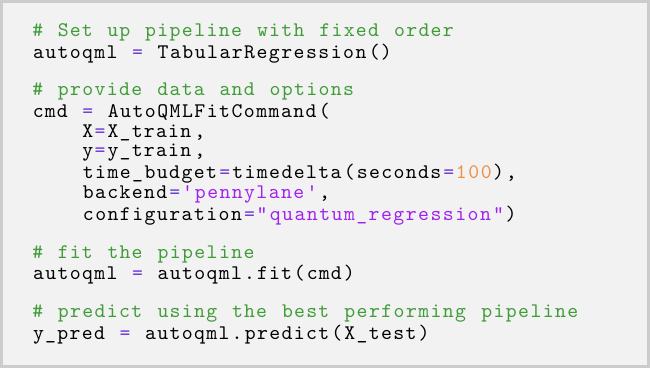}
    \caption{Example code for fitting a tabular regression pipeline. Here, it is assumed that the training data is supplied as \texttt{X\_train} with corresponding targets \texttt{y\_train}. Within AutoQML, the data is split into a test and validation set. Options such as the time budget \texttt{timedelta} for the optimization or the backend for execution of the QML algorithms can be specified. In the example, the preset configuration \texttt{"quantum\_regression"} is used to restrict the search space to quantum computing based regression algorithms only.}
    \label{fig:code-example}
\end{figure}

To solve Eq.~\eqref{eq:optimal_pipeline} AutoQML utilizes probabilistic models to guide the search for high-performing configurations. It leverages Bayesian optimization using tree-structured Parzen Estimator (TPE)~\cite{NIPS2011_86e8f7ab}. 
The search algorithm constructs a Gaussian mixture model to approximate the loss (e.g., validation loss or accuracy) based on the results of previous evaluations. For a given ML problem, the optimizer performs the following steps: (i) For a new pipeline \(i\), a configuration $\vec{\lambda}_i$ is drawn from \(\Lambda\) by sampling close to the optimal point of a probabilistic model, which favors promising regions of the search space. (ii) The proposed configuration $\vec{\lambda}_i$ is passed to the evaluation function, yielding a performance score $l_i$. This score reflects how well the pipeline performs on a validation set. (iii) The tuple $(\vec{\lambda}_i, l_i)$ is used to update the probabilistic model of the loss function. The model is refined to more accurately represent the relationship between configurations and their performance. (iv) The process repeats until a user-provided budget, such as a time limit, is exhausted. When the optimization process is finished, the best-performing configuration \(\vec{\lambda}^\star\) and the corresponding fitted pipeline are returned to the user. This optimization process enables AutoQML to efficiently discover high-performing (quantum) ML pipelines tailored to the specific problem at hand. A code example is shown in Fig.~\ref{fig:code-example}.

The fixed pipeline templates in AutoQML account for quantum computing specific preprocessing. Particularly, the limited size of current quantum computer requires a significant dimensionalty reduction, e.g.,  through principal component analysis (PCA) or autoencoders. Additionally, the encoding of classical data in AutoQML is done using pre-defined encoding circuits based on angle-encoding. This usually requires scaling the features to avoid non-injective maps.

\subsection{QML Integration}
\label{sec:squlearn}

The QML algorithms are provided by sQUlearn. The library offers several high-level methods such as quantum neural networks (QNN)~\cite{PhysRevA.98.032309}, quantum reservoir computing (QRC)~\cite{Innocenti2023}, and various kernel methods such as quantum kernel ridge regression (QKRR)~\cite{schnabel2024quantumkernelmethodsscrutiny}, quantum support vector machines (QSVM)~\cite{Havlicek2019}, and quantum Gaussian processes (QGPR)~\cite{Rapp2024}. 
These methods are accessible through a scikit-learn programming interface, allowing for user-friendly implementation. High-level methods can be modularly configured using a variety of pre-implemented quantum encoding circuits. Each circuit's configuration can be tailored by adjusting hyperparameters such as the number of qubits and the number of layer repetitions.
In QNNs and QRC, the form and quantity of observables used to compute the output can also be customized by hyperparameters. Additionally, fidelity kernels and projected quantum kernels are available for quantum kernel methods. Together with classical hyperparameters, such as regularization strength for the QSVM or optimization parameters for QNNs, these degrees of freedom form the configuration space for the QML predictors.

The QML methods can be executed using PennyLane and Qiskit simulators, as well as IBM Quantum computers and several quantum computing backends provided by Amazon Braket. In the case where real quantum computers are used for execution, communication with the quantum hardware providers is managed by sQUlearn. For IBM quantum backends, an automated hardware selection routine that can prioritize either speed or accuracy is available.

\section{Use Cases}
\label{sec:use_cases}
We benchmark AutoQML on four scenarios based  on real-world use cases from the domains of manufacturing and automotive. For each, we derive a supervised learning problem. The learning problems have been chosen such that their (effective) dimensionality and data set size are small enough so that they can be processed by current quantum computers (or simulators) while still retaining an adequate level of difficulty. For some use cases this required creating synthetic data.

In the following, we briefly describe each use case and the data sets used to benchmark the framework. For some problems, the data sets are preprocessed before being inputted into AutoQML. In these cases, the preprocessing steps are described in the corresponding section for each use case.

\subsection{Time Series Classification}
\label{sec:time_series_classification}
We consider a time series classification task in which sensor data is collected from an autonomous vehicle to identify several states of the vehicle, for example, the execution of individual tasks or abnormal activity. To this end, we synthetically generate vibration sensor data, i.e., a univariate time series containing $7$ individual states which are each associated with a unique label for the classification task. From the time series, we generate a spectrogram that is then divided into 2-dimensional tiles of size $\left[ 7\times 30\right]$. Each tile is flattend into a one-dimensional vector of dimension $d=210$ and a label corresponding to the states present in the tile is assigned. In total, the data set contains $N=3291$ samples that are divided into $N_{\text{training}} = 758$ training points $N_{\text{test}} = 2\, 533$ testing points. Since most tiles are associated with no sensor activity, the training set is stratified so that all classes occur roughly with the same frequency.

\subsection{Image Classification}
\label{sec:image_classification}
In sheet metal processing using laser cutting machines, metal plates rest on a bed of supporting slats during processing. These slats are manually positioned by the machine operator in fixed socket positions ahead of time, with the specific configuration depending on the cutting task. For optimal operation, it is beneficial to know the positions of the slats in advance~\cite{STRUCKMEIER2019575}. The associated task is a binary image classification problem, where the goal is to determine whether a given image contains a supporting slat.
To ensure enough image data for training, we use a proprietary synthetic pipeline to automatically generate artificial yet authentic slat images using accurate CAD models of all employed parts a 3D rendering framework. These rendered images feature entirely randomized configurations of slat positions. Since each rendered image depicts multiple slats, it is subsequently divided into smaller image snippets of size $80 \times 200$ pixels, each focusing on exactly one slat position, which is either occupied or not.
Although this methodology allows for the generation of an arbitrary number of images, to keep computational requirements moderate, we restrict the data set under study to 500 image snippets, divided into $N_{\text{training}}=400$ training and testing points $N_{\text{test}}=100$. Moreover, we perform feature reduction via PCA in advance, reducing each image snippet to $d = 8$ features. This reduction is necessary to maintain the confidentiality requirements of the use case. We treat the resulting latent space vectors as input images. The data set has been presented in more detail by~\citet{basilewitsch2024quantumneuralnetworkspractice}.

\begin{figure*}[t]
    \centering
    \includegraphics[width=0.85\linewidth]{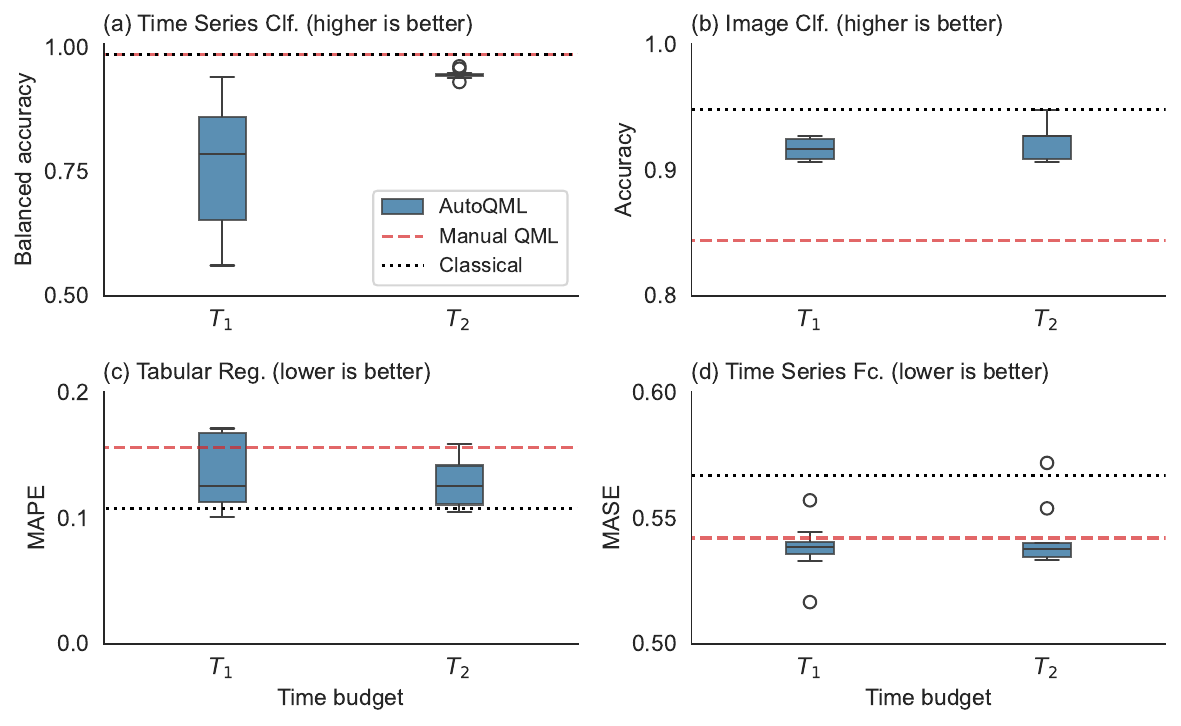}
    \caption{Performance of AutoQML (boxes) for two different time budgets $T_1$ and $T_2$. Additionally, manual QML pipelines (red, dashed) and classical models (black, dotted) are depicted. (a) shows the balanced accuracy (higher is better) for the time series classification. (b) shows the accuracy (higher is better) for the image classification. (c) shows the mean absolute percentage error (MAPE, lower is better) for the tabular regression, and (d) shows the mean absolute scaled error (MASE, lower is better) for the time series forecasting. For the box plots, points that are outside $1.5\times$ the inter-quartile range are shown as circles, and the lines inside the boxes denote the sample median.}
    \label{fig:autoqml-results}
\end{figure*}

\begin{figure*}[t]
    \centering
    \includegraphics[width=\linewidth]{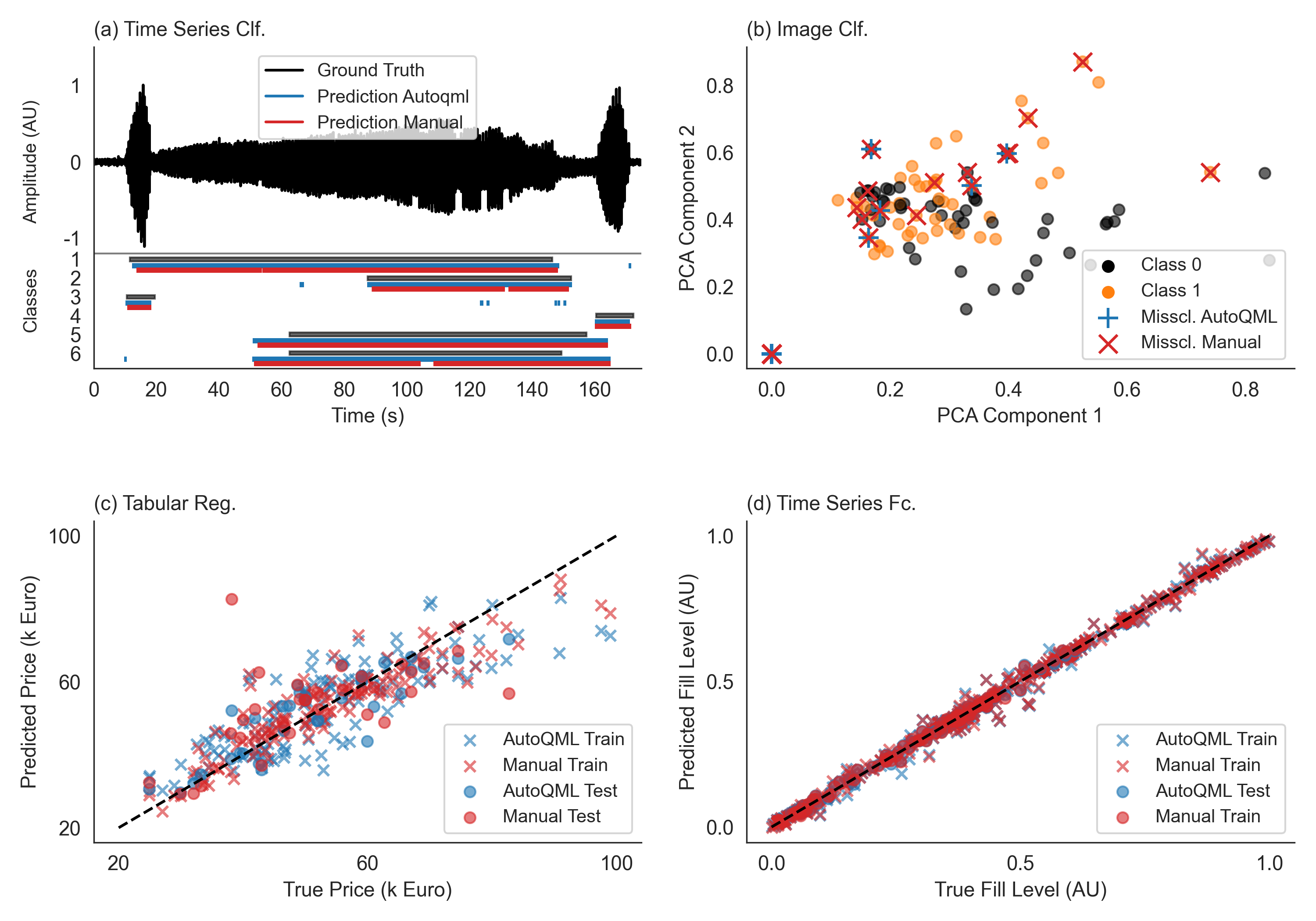}
    \caption{Application of the best AutoQML pipelines (blue) from Fig~\ref{fig:autoqml-results} on the respective use cases. The application of the manually created models is shown in red. (a) shows signal in the upper part of the figure. The bars bellow show the presence or absence of events which are classified by the models. (b) shows the two principal components with the larges singular values of the test set of the image classification use case. The points that have been missclassified by the AutoQML (cross, blue) or the manual pipeline (plus, red) are shown in addition. The classes 0 (no slat) and 1 (slat) are shown in different colors. Figure (c) and (d) depicts the prediction vs. the true target values of the tabular regression and time series forecasting, respectively.
    \label{fig:autoqml-use-cases}}
\end{figure*}

\subsection{Tabular Regression}
\label{sec:tabular_regression}
Accurate price forecasting is essential for companies managing pre-owned assets, whose values fluctuate with spatial and temporal variations in supply and demand. This is particularly relevant for heavy construction equipment dealers and rental companies, who depend on precise price predictions to optimize asset management. Assessing the current and future residual value of their fleets enables these companies to determine the ideal time to resell individual pieces of machinery. By collecting data from seven major online construction equipment portals, we create a data set with $N=165$ data points ($N_{\text{training}}=132$ and $N_{\text{test}} = 33$). Each data point represents one \emph{Caterpillar} type 308 construction machine. The features are the construction year, the working hours, the current location and the model extension. The target values are the prices. The location and model extension are categorical variables with 9 and 3 unique values, respectively. The resulting data set has six dimensions (16 if one-hot-encoded). The data set is a subset of the data used by~\citet{stuhler2024evaluating}.

\subsection{Time Series Forecasting}
\label{sec:time_series_forecasting}
Engineering control technology systems for the automotive sector heavily depend on the ability to model or simulate sensor time series data. This is particularly relevant for dynamic situations, such as when accelerating a vehicle. Producing accurate and precise forecasts of physical quantities can significantly influence the quality of system control. To manipulate, compare, and analyze the computed models effectively, we select the relative cylinder filling of an internal combustion engine as an application where we can easily control the dimensionality. The time series encodes complex non-linear dynamics.
Using a sliding-window approach, the problem is formulated as a regression task, and the number of time steps utilized for the forecast can be freely chosen. Our real industrial data set consists of $10\, 000$ time steps, each covering a period of $10\,\text{ms}$. The final data set is a resampled version with $N_{\rm train}=556$ and $N_{\rm test}=140$, where the features are lagged versions of the time series from the four previous time steps. Training and test data are derived from different parts of the time series. 

\section{Results}
\label{sec:results}

The performance of the AutoQML framework is evaluated using the four use cases described in Sec.~\ref{sec:use_cases}. The results are shown in Fig.~\ref{fig:autoqml-results}. For each use case, we fit AutoQML for ten different seeds with two time budgets, $T_1=\autoqmlshort\, \mathrm{s}$ and $T_2=\autoqmllong\, \mathrm{s}$. Note that although AutoQML can optimize over a joint algorithm pool consisting of classical and quantum ML algorithms, we are interested in the performance of the QML algorithms in particular and thus only include quantum methods in the search space for this benchmark.

We compare our results with manually created QML pipelines (red, dashed lines). 
For two use cases, we sourced manual solutions from previous studies~\cite{stuhler2024evaluating, basilewitsch2024quantumneuralnetworkspractice}, while for the other two, we constructed custom pipelines tailored to the specific use cases. Details of the manual models and the process used to obtain them are provided in Appendix~\ref{sec:manual_models}. Since these models have been crafted by quantum computing specialists, they require significantly more expertise and time compared to the corresponding AutoQML solutions. All quantum models are evaluated using the PennyLane statevector simulator. Additionally, to better gauge the quality of the results, we compare them with the performance of classical models. These are shown as dotted lines in Fig.~\ref{fig:autoqml-results}, indicating the performance of the best model among random forests, XGBoost~\cite{10.1145/2939672.2939785}, and support vector machines. For regression, Gaussian process regression is also included. For the kernel methods, RBF kernels were used, and the hyperparameters of all models were optimized using Optuna. The preprocessing pipeline is the same as for the manually created QML models.

When evaluating the manually obtained QML pipelines against the AutoQML pipelines, we observe that the performance is comparable. For three out of the four use cases, the AutoQML pipeline outperforms the manual QML pipeline on average, with a slight improvement in the tabular regression and time series forecasting use cases (b, c) and a clear advantage in the image classification use case (d). However, for the time series classification use case (a), the manual pipeline provides a superior solution.

In all use cases, we observe a median improvement when granting a larger time budget. Specifically, for the time series classification use case (a), there is a significant performance increase with the budget $T_2$ compared to $T_1$. However, for the other use cases, the performance gains are only marginal. Additionally, the drastically reduced variance with time budgets $T_2$ suggests that budget the $T_1$ is insufficient for the time series classification use case, while for other use cases, the pipeline search appears to be nearing convergence even with $T_1$. This difference in convergence is likely due to the much larger dimensionality of the time series classification use case compared to the other use cases.

\begin{table*}[t]
    \centering
    \caption{Summary of different model pipelines. Fidelity quantum kernels are denoted by FQK. The row \emph{Observables} depicts the measurement observables for projected quantum kernels (PQK) or QNNs. Here, $X_i,Y_i,Z_i$ denotes the respective Pauli operator on qubit $i$ where $i=1,\dots,n$ runs across all qubits. The description of the outer kernels in cases where PQKs have been used (in the present cases pairwise, Matter, dot product), and the encoding circuits can be found in the sQUlearn documentation~\cite{squlearn-docu}.}
    \begin{tabular}{l@{\hskip 10pt}l@{\hskip 10pt}l@{\hskip 10pt}l@{\hskip 10pt}l@{\hskip 10pt}}
    \toprule\toprule
    & Time Series Classification & Image Classification & Tabular Regression  & Time Series Regression \\     
    \midrule
    Dim. Red. & PCA & UMAP & -- & -- \\ 
    Scaling & Normalization & Standardization & Normalization & Normalization \\ 
    Observable & -- & $\lbrace X_i, Z_i\rbrace$ & $\lbrace X_i, Y_i, Z_i\rbrace$ & $\lbrace X_i, Y_i\rbrace$ \\ 
    Model & QSVM; FQK & QSVM; PQK(Mattern) & QKRR; PQK(DotPorduct) & QGPR; PQK(pairwise) \\ 
    Encoding Circuit & \cite{Hubreg_} & Multi-Control & Multi-Control & YZ-CX \cite{Haug_2023} \\ 
    Num. Qubits & 8 & 8 & 8 & 8 \\ 
    Num. Layers & 1 & 2 & 3 & 3 \\
    \bottomrule\bottomrule
\end{tabular}
    \label{tab:autoqml_pipeline}
\end{table*}

When comparing the QML models with the classical models, we observe that the best AutoQML pipeline outperforms the best classical solution in three out of the four use cases (b–d), while for the time series classification (a), the classical solution achieves the highest score. Overall, the classical solutions are similar to the AutoQML solutions, and neither demonstrates a clear advantage over the other. The comparable quality of results across all models supports the validity of the manual QML models and provides evidence for the effectiveness of AutoQML.

The pipelines with the highest scores in Fig.~\ref{fig:autoqml-results} are shown in Tab.~\ref{tab:autoqml_pipeline}. The rows show the choices for the corresponding pipeline step. Some preprocessing steps such as one hot encoding are only relevant for specific use cases are omitted in the table. The final pipelines are diverse in terms of models and preprocessing. Notably, all models are quantum kernel models. This is, at least partially, expected since training QNNs is significantly more time consuming than training quantum kernel methods. Therefore, QNNs are underrepresented in the optimization process. Estimators based on QRC were the best optimizers in several runs. However, in none of the use cases was QRC the best among the ten runs (i.e., the ten optimization with different seeds per use case). Three out of the four kernel methods use projected quantum kernels (PQK). Interestingly, all PQKs in the best performing models do not employ the commonly used RBF-type outer kernel function. This is in agreement with other studies which found that the outer kernel function in PQKs should be treated as an additional hyper-parameter~\cite{schnabel2024quantumkernelmethodsscrutiny}.

Figure~\ref{fig:autoqml-use-cases} shows the application of the best AutoQML pipelines (blue) from Fig.~\ref{fig:autoqml-results} together with the models from the manually crafted pipelines. Overall, the pipelines solve the use cases well. In the time series classification (a), both pipelines are able to determine the classes, although classes 4 and 5 seem to be more difficult as both pipelines do not classify them optimally. In the image classification (b), the pipelines are applied to the full data set but only the first to principal components are shown for visualization. It can be seen that the first two principal components are not sufficient to linearly separate the model. This aligns with the result of the best AutoQML pipeline, which uses all 8 principal components present in the data set. Most of the misclassifications happen in the region where the classes overlap in the first two components, indicating that even with the full dimensionality, there might still be some overlap of the data classes. The application of the pipelines to the regression problems (c) and (d) is in line with the expectations from the performance metrics in Fig.~\ref{fig:autoqml-results}. Overall, the tabular regression use case (c) is more difficult than the time series forecasting problem because the tabular regression problem has higher dimensionality and more noise compared to the relatively simple one-step ahead problem in (d). This difference in difficulty is reflected in the larger deviations of the predicted values from the true values in (c) compared to (d). Comparing the training performance to the test performance, no significant shortcomings, i.e., overfitting or underfitting, can be observed.

\section{Discussion}
\label{sec:discussion}

The results in this paper have been performed with statevector simulations. As quantum computing matures, a signficant portion of the model evaluation will have to be done on quantum computers. Although we have tested this prototypically, a full pipeline optimization on current quantum hardware is currently infeasible due to both financial and time constraints. Since simulation techniques co-evolve with the hardware~\cite{10.1145/3564246.3585234, angrisani2024classicallyestimatingobservablesnoiseless}, we foresee that in the upcoming years the pipeline search will most likely involve a combination of advanced simulation techniques and evaluation on real devices with a requirement to be efficient in QPU time as much as possible.
Furthermore, the current evaluation time for QML models is notably longer than for classical ML models, necessitating a greater time budget for pipeline optimization. This underscores a trade-off between human developer time and computational time, which is more significant than in classical AutoML frameworks. Addressing these challenges and integrating novel QML developments into AutoQML will be crucial for advancing the framework's efficacy and versatility.

Although we have tested AutoQML on a diverse set of problems, the framework in its current form is only designed for supervised ML problems. Extensions to unsupervised problems like clustering are conceivable. Through its modular design and the encapsulation of the quantum computing-facing modules in sQUlearn, such extensions can be implemented easily. This is also true for incorporating novel developments in QML, such as new models.
    
Currently, AutoQML only supports a fixed, predefined library of circuits. These circuits can be further customized by the automation process. Nevertheless, recent work indicates that tailoring QML models to the dataset, rather than relying on generic hardware-efficient circuits, might be required to retain trainability as the models grow in depth and width~\cite{NEURIPS2021_69adc1e1, Cerezo2022}. Incorporating automated approaches to quantum circuit design~\cite{Rapp_2025} into the framework is thus left for future work.

\section{Conclusion}
\label{sec:conclusion}

We have introduced AutoQML, an innovative framework for automated QML, and evaluated its performance through comprehensive benchmarking on a diverse set of problems derived from four distinct industrial applications, specifically two classification and two regression tasks. Our results demonstrate that AutoQML is capable of effectively generating QML pipelines that incorporate QML-specific preprocessing, model selection, and hyperparameter optimization. Notably, the performance of the generated pipelines was competitive with that of manually constructed ones, which were, wherever feasible, derived from existing literature to reduce subjective biases. These findings indicate that AutoQML is a valuable tool for addressing machine learning challenges in QML, requiring minimal expertise in quantum computing. Additionally, the capabilities of AutoQML underscore its potential as a powerful prototyping and benchmarking resource for QML researchers and practitioners.

\begin{acknowledgements}
This work was supported by the German Federal Ministry
of Economic Affairs and Climate Action through the
project AutoQML (grant no. 01MQ22002A). We thank Frederic Rapp, Timoteo Lee and Khaled Al-Gumaei for their contributions to the time series classification use case. We thank Peter Schichtel and Jaroslav Vondrejc for their contribution to the time series forecasting use case. We thank Giorgio Silvi for the implementation of the automated backend selection and Dennis Kleinhans for the implementation of tests for the framework. The authors disclose the use of LLM-based tools for grammar and spelling.
\end{acknowledgements}

\appendix
\section{Manual QML Models}
\label{sec:manual_models}
In this section, we briefly describe the pipelines for the manual solutions and the process by which they were obtained.

\subsection{Time Series Classification}
To reduce the dimensionality of the data set, we perform a PCA with 5 components and scale the output to the interval $[-1,1]$. The manual model is a QSVM using a projected quantum kernel with the feature map from~\cite{Hubreg_} with 5 qubits and 6 layers. The model is obtained using hyperparameter optimization over the regularization parameters of the QSVM, as well as the number of layers and qubits. The model used in the benchmark is the best performing from optimizing over a set of two quantum feature maps~\cite{Hubreg_, kreplin2023reduction}.

\subsection{Image Classification}
The manually created pipeline contains no preprocesing steps in addition to those described in Sec.~\ref{sec:image_classification}. The model is a QNN classifier with 8 qubits and an Ising-type cost operator. The model is the best performing model presented by~\citet{basilewitsch2024quantumneuralnetworkspractice}. Details on how the circuit has been determined can be found there.

\subsection{Tabular Regression}
The reference quantum model is obtained from Ref.~\cite{stuhler2024evaluating}, in which it demonstrates optimal performance on a similar data set for a different type of construction machinery. In line with the original study, categorical features are one-hot encoded, resulting in a dataset with a dimensionality of $d=15$. These features are then scaled to the interval $[-1, 1]$. The reference model in Fig.~\ref{fig:autoqml-results} is a QSVM that utilizes a Fidelity Quantum Kernel. The encoding circuit has been obtained from Fig. 5 of Ref.~\cite{stuhler2024evaluating}, and employs 15 qubits (one qubit per feature). The hyperparameters of the underlying SVM have been optimized through a dedicated hyperparameter optimization process.

\subsection{Time Series Forecasting}
The manually created pipeline contains no additional pre-procesing steps. The model is a 4-qubit quantum reservoir regressor with $54$ random measurement operators which are fed into a linear regression model. The used reservoir is the result of a search over the number of qubits, the number of layers used in the encoding, the number of observable,and the architecture of the encoding circuit. The search is performed using Optuna.

\bibliography{autoqml}

\end{document}